\documentclass[aps,prl,twocolumn,groupedaddress]{revtex4}

\usepackage{graphics} 
\usepackage{graphicx}
\usepackage{dcolumn}
\usepackage{bm}
\usepackage{epstopdf}

\begin{document}

\title{Shockley-Queisser Model: Analytical Solution, Thermodynamics, and Kinetics}

\author{Andrei Sergeev}
\email[Corresponding author: ]{podolsk37@gmail.com}
\author{Kimberly Sablon}
\affiliation{U.S. Army Research Laboratory, Adelphi, MD 20783, USA}

\date{\today}

\begin{abstract} 

We derive exact analytical solution of the Shockley-Queisser (SQ) model and present all photovoltaic (PV) characteristics in compact and convenient form via the Lambert W function. We show that the SQ condition of chemical equilibrium between photocarriers and emitted photons leads to a new thermodynamic relation between the maximal conversion efficiency ($\eta_m$) and the photo-induced chemical potential ($\mu_m$):  $ \eta_m =(\mu_m-kT) / \epsilon^*$, where $\epsilon^*$ is the average energy in the incoming photon flux per absorbed photon and  $\mu_m$ is the corresponding chemical potential of electrons and photons, i.e. $\mu_m /q= V_m$  is the output voltage at optimal conversion ($q$ is the electron charge). We consider kinetics of PV conversion and calculate the optimal photocarrier collection time. The above formalism is applied to semiconductor materials to specify photonic and electronic requirements for approaching the SQ limit.

\end{abstract}
  

\maketitle

In 1960 Shockley and Queisser published their famous paper \cite{1}, where they developed universal and fundamental model of PV conversion in a semiconductor single-junction solar cell and numerically calculated the maximal conversion efficiencies as function of spectral properties, radiation power (light concentration), and a semiconductor bandgap. The Shockley-Queisser (SQ) limiting efficiencies are considered as the most fundamental benchmarks in solar light conversion. Despite hundreds of works devoted to the SQ model, a number of important questions, -interrelation of the SQ limit with classical thermodynamics \cite{2}, endorewersible processes \cite{3}, nonequilium thermodynamics \cite{4}, and a possibility to overcome the SQ limit due to nanoscale photonic management \cite{5,6} and nano-enhanced thermophotovoltaic conversion \cite{7}, - are still hot topics of modern research. In this work we derive analytical solution of SQ model and present main PV characteristics in simple and convenient forms. Using this solution we establish new general relations between PV characteristics. 

The S-Q model is based on the three-stage photocarrier kinetics in a semiconductor. At the first stage, the light-induced carriers lose energy due to strong electron-phonon interaction. As a result, the photocarriers acquire equilibrium temperature and relax to the band edge. At the second stage the photocarriers emit photons that are reabsorbed and create the electron-hole pairs again. The reabsorption process establishes the chemical equilibrium between photocarriers and emitted photons. In formal words, the photocarriers and emitted photons are describes by the same temperature $T$ and chemical potential  $\mu$.  Thus, the corresponding light-induced distribution functions of photocarriers and emitted photons are $f_e = [\exp(\epsilon - \mu)/kT +1]^{-1}$   and $f_{ph} = [\exp(h\omega - \mu)/kT -1]^{-1}$  with the chemical potential $\mu$ independent on $\epsilon$ and $\omega$.  At the third stage, a stationary value of the light-induced chemical potential is estalished due to carrier collection at the device contacts and due to photon escape from the device. The stationary photon and photocarrier distribution functions, may be approximated by the quasi-classical distribution, $f_{ph}=\exp[(\mu-h\omega)/kT]$ and $f_e =\exp[(\mu-\epsilon)/kT]$, because the light induced chemical potential is still substantially below the bandgap and the parameter $(\epsilon-\mu)/kT$  is much larger than 1 even in the case of the maximal solar light concentration. The quasi-classical function may be factorized as $\exp(\mu/kT)\times \exp(-h\omega/kT)$  and, therefore, the flux of the emitted photons may be presented as $\dot{N}_{em}(\mu,T)= \exp(\mu/kT)\cdot \dot{ N}_{em}(T)$, where $ \dot{N}_{em} (T)$  is the emission flux in thermodynamic equilibrium ($\mu=0$). If nonradiative recombination is negligible, the electric current is determined by the difference between absorbed and emitted photon fluxes, $J/q= \dot{N}_{ab} - \dot{N}_{em}$. Taking into account that the useful energy per electron is $\mu$, the photovoltaic conversion efficiency is given by
\begin{eqnarray} \nonumber
\eta &=&  {\mu (J / q) \over \dot{E}_{in} } = \left(  N_{ab} -N_{em}(T) \cdot \exp \left(  {\mu \over kT} \right)  \right) \cdot {\mu   \over  \dot{E}_{in}  }  \\
&=& \left(  1- { \dot{N}_{em} (T) \over \dot{N}_{ab} }  \cdot \exp \left(  {\mu \over kT} \right)  \right)   \cdot {\mu \over \epsilon^*},
\label{1}
\end{eqnarray}
where $ \dot{E}_{in}$  is the power of incoming photon flux and $ \epsilon^*=  \dot{E}_{in} / \dot{ N}_{ab}$ is the average energy in the flux per a photon absorbed. As an example, Fig. 1 shows the parameter  $ \epsilon^*$ as a function of the bandgap for the 6000 K thermal radiation and 100 \% absorption above the bandgap. 

\begin{figure}[t]
\includegraphics[width=3.3in]{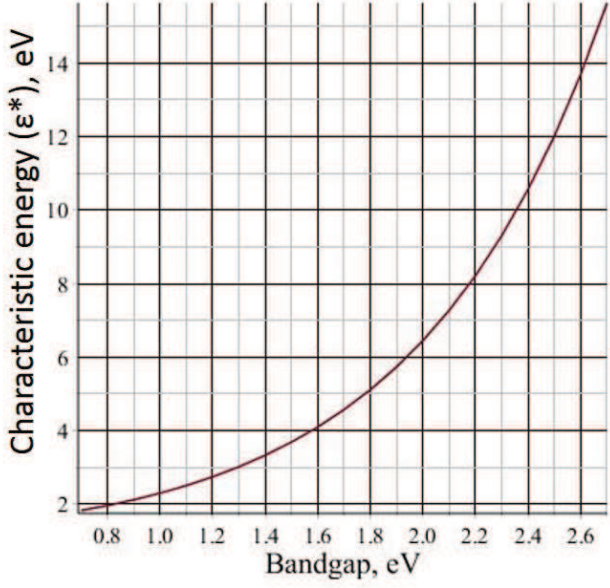}
\caption{Average energy per absorbed photon, $\epsilon^*$, for the 6000 K radiation flux as a function of semiconductor bangap (100\% absorption above the bandgap).}
\label{epssilonstar}
\end{figure}

Introducing the dimensionless parameter, 
\begin{equation}
A= { \dot{N}_{ab} \over \dot{N}_{em} (T)},
\label{2}
\end{equation}
and by optimizing the efficiency as a function of the chemical potential (Eq. 1), we obtain the following equation for the chemical potential (output voltage) at maximal PV efficiency,
\begin{equation}
\left( {\mu_m \over kT} +1 \right) \cdot \exp{\mu_m \over kT}=A
\label{3}
\end{equation}
This transcendental equation has analytical solution in terms of the Lambert W function, 
\begin{equation}
{\mu_m \over kT} = {\rm LW} (A\cdot e) -1,
\label{4}
\end{equation}
where LW  is used for the Lambert W function and $e$ is the base of natural logarithm.
Substituting $\mu_m$ to the equation for the conversion efficiency (Eq. 1), we find the maximal conversion efficiency,
\begin{equation}
\eta_m = \left[  {\rm LW} (A\cdot e) -2 + {1\over {\rm LW} (A\cdot e)} \right] \cdot {kT \over \epsilon^*}  .
\label{5}
\end{equation}
For $A \gg 1$, we can use asymptotic formula for the Lambert W function, 
\begin{equation}
{\rm LW}(z) = \ln(z) - \ln\ln(z) + {\ln\ln(z) \over \ln(z)}, \ z \gg 1  .
\label{6}
\end{equation}
Taking into account the asymptotic expression for the Lambert W function and neglecting the terms of the order of $1/\ln(A)$, we can simplify Eqs. 4 and 5. With $1/\ln(A)$  accuracy, we get
\begin{eqnarray}
qV_m &=& \mu_m=kT \cdot {\rm LW} (A),  \\
\eta_m &=& \left[{\rm LW}(A) -1\right]\cdot {kT \over \epsilon^*}.
\label{78}
\end{eqnarray}
Finally, using Eqs. 7 and 8 we obtained the general relation between conversion efficiency and $ V_m$,
\begin{equation}
 \eta_m = {qV_m \over \epsilon^*} \left( 1 - {kT \over qV_m} \right) = {qV_m-kT  \over \epsilon^*}  .
\label{9}
\end{equation}
As it is expected, at T=0 the conversion efficiency is $qV_m/\epsilon^*$, i.e. the ratio of the useful energy of photoelectron to the average energy in the incoming flux per absorbed photon. Eq. 9 establishes a new thermodynamic relation, which shows that in the PV conversion the photon emission reduces the useful energy of a solar photon by $kT$ independently of the bandagap of the PV material, spectral properties and power of the converted radiation.  It is important to note that while the direct emission losses are of the order of $ kT/\epsilon^*$ and reduces the PV efficiency just by 0.5 -  2 \%, the photon emission plays a critical role because the nonequilibrium chemical potential and, therefore, $V_m$ are also determined by the photon emission.

It is convenient to express the solar cell performance via the open circuit voltage, 
\begin{equation}
 qV_{oc} = \mu_{oc} = kT \cdot \ln(A).
\label{10}
\end{equation}
The voltage at maximum efficiency is the universal function of $V_{oc}$ and may be presented as
\begin{eqnarray} \nonumber
 && V_M = {kT \over q} \cdot {\rm LW} \left(  \exp \left(  {qV_{oc} \over kT} \right) \right)  \\
&& \approx V_{oc} - {kT \over q } \cdot \ln \left( {q V_{oc} \over kT} \right)+ {(kT)^2 \over q^2 V_{oc}} \cdot  \ln \left( {q V_{oc} \over kT} \right) 
\label{11}
\end{eqnarray}
Solution of the SQ model for the PV efficiency (Eq. 8) also provides a simple formula for the fill factor, 
\begin{eqnarray} \nonumber
&&  FF \equiv {\eta \dot{E}_{in} \over J_{sc}  \cdot V_{oc} }= { {\rm LW} (A) -1 \over \ln(A) }  \\
&& \approx 1 -{1 +  \ln ( q V_{oc} / kT)  \over  q V_{oc} / kT}  +  {\ln ( q V_{oc} / kT) \over ( q V_{oc} / kT)^2} .
\label{12}
\end{eqnarray}

The nonradiative recombination processes may be easily incorporated in the formalism above in terms of quantum efficiency, $\gamma$ \cite{9}.  In this description, the electric current is given by $J/q = \gamma \cdot \dot{N}_{ab} - \dot{N}_{em}(\mu, T)$. Using Eqs. 1-5 we obtained the conversion efficiency (see Eq. 8), 
\begin{equation}
 \eta_m =  [{\rm LW} ( A_\gamma) -1]  \cdot { kT \over \epsilon^*_\gamma},
\label{13}
\end{equation}
where $ A_\gamma = \gamma \cdot A$  and   $\epsilon_\gamma^* = \gamma \cdot  \epsilon^*$. Spectral dependence of quantum efficiency, $\gamma(\lambda)$,  can be also incorporated in the above formalism via corresponding integrals for $ A_\gamma$  and $\epsilon_\gamma^*$. In Eq. 13 the quantum efficiency can exceed 100\% due to multiple exciton generation, which is observed in quantum dot PV materials \cite{8}.  

Eq. 13 assumes that quantum efficiency does not depend on chemical potential or that the QE was determined at maximal conversion efficiency. The Shockley-Read-Hall (SRH) recombination is proportional to $\exp(\mu/2kT)$ and the corresponding model does not allow analytical solution. Effects of SRH processes can be calculated in perturbative way.  Let us define $ k$ as a ratio of the SRH recombination rate to the radiative rate in equilibrium ($\mu = 0$), i.e. the ratio of SRH and radiative dark currents, $J_{02}/ J_{01}$, in the double diode model.  In the first order in $k$, the decrease of the voltage at optimal conversion is given by 
\begin{equation}
 {\Delta_{SRH} V_m \over V_m} = k {1 \over 2\sqrt{A\cdot{\rm LW} (A)}}.
\label{14}
\end{equation}
  
As the Lambert W function is a built-in function in all popular mathematical software, Eqs. 5 and 13 allows one directly calculate PV efficiency without any programing and optimization at the J-V curve. Fig. 2 shows the results of such calculations for concentrated and unconcentrated 6000 K radiation. 

\begin{figure}[t]
\includegraphics[width=3.3in]{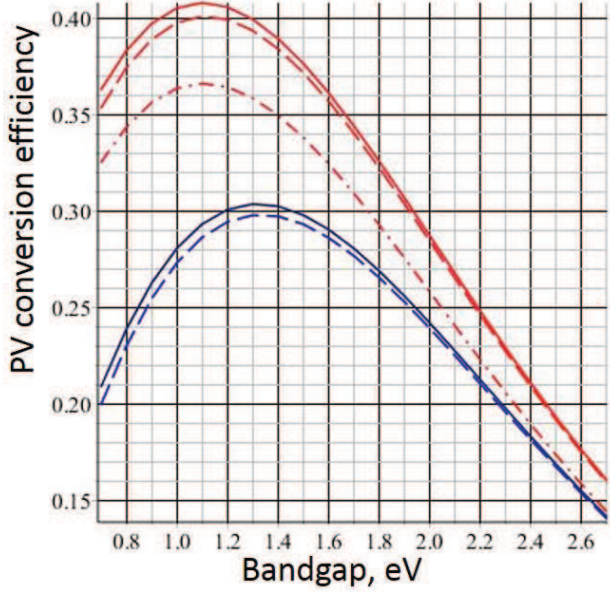}
\caption{PV conversion efficiencies given by Eq. 8 for the concentrated light and backward mirror (solid red), without mirror (dashed red), 90 \% QE (dashed-dotted red), unconcentrated light with backward mirror (solid blue), and without mirror (dashed blue).}
\label{efficiency}
\end{figure}

Next, we will employ the developed formalism to kinetics of photons and electrons in a semiconductor film and calculate corresponding characteristics times and lengths. The number of thermally excited photon modes in the film of thickness $d$ is $ N(T,d) = d\cdot \int d\omega \ D_{ph}(\omega) f_{ph} (\omega) $, where$ D_{ph}(\omega)$  is the photon density of states and $f_{ph}(\omega)$ is the Plank distribution function. The emission of these modes from the film is limited by internal reflection at the film-air interface and the emission rate is given by $ N(T,d) =  (\tilde{c} /4) (\cos \theta_{tr})^2 \cdot \int d \omega \ D_{ph} (\omega) \ f_{ph} (\omega) $ , where $\theta_{tr}$  is the angle of total internal reflection, ($\cos \theta_{tr} =1/n$), and $\tilde{c} $ is the light velocity in the film. Therefore, the photon time of escape from the film is given by 
\begin{equation}
\tau_{es} = {N(T,d) \over \dot{N}_{em} (T)} = 4n^2 {d \over \tilde{c}}.
\label{15}
\end{equation}

 Isotropic photon distribution assumed in SQ model delays the photon escape by the factor $4n^2$ introduced by Yablonovich \cite{10}. As we discussed above, the photon reabsorption establishes chemical equilibrium between photoelectrons and the emitted photons. In this limit the photon escape time is much longer than the photon absorption time, $\tau_{ab} = 1/(\alpha \tilde{c})$ , where $\alpha^{-1} $ is the absorption length of photons with energy slightly above the bandgap. The reabsorption increases the photocarrier lifetime, which is given by
\begin{equation}
\tau_\ell = \tau_R {\tau_{es} \over \tau_{ab}} = 4n^2 d \alpha \cdot \tau_R
\label{16}
\end{equation}
 where $\tau_R$   is the minority carrier recombination time with respect to photon emission. To calculate the optimal photocarrier collection rate, let us present the electric current and the photon emission rate in terms of the carrier collection time, $\tau_{col}$, and photoelectron lifetime $\tau_\ell$ \cite{11},
\begin{eqnarray} \nonumber
 J(V) =  {q n_0 d \over \tau_{col}(V) } \cdot \exp{ qV \over kT}, \\
\dot{N}_{em} (V) = {n_0 d \over \tau_\ell} \cdot \exp{ qV \over kT},
\label{17}
\end{eqnarray}
where $n_0$  is the equilibrium concentration of carriers. Taking into account that $J(V)/q = \dot{N}_{ab}-  \dot{N}_{em}$ and using Eq. 7, we find that the photocarrier collection time in the optimal regime is given by
\begin{eqnarray} \nonumber
 \tau_{col}(V_m) = \tau_\ell / \beta, \\
\beta = {\rm LW} (A) = {qV_m \over kT}.
\label{18}
\end{eqnarray}

Fig. 3 shows the dependence of parameter $\beta$ on the semiconductor bandgap for the 6000 K radiation. As seen, in the optimal conversion regime the carrier collection time is substantially shorter than the photocarrier lifetime. For example, for PV conversion of unconcentrated light with Si solar cell $\ beta$ equals 29 and with GaAs solar cell $\beta$ equals 40.

\begin{figure}[t]
\includegraphics[width=3.3in]{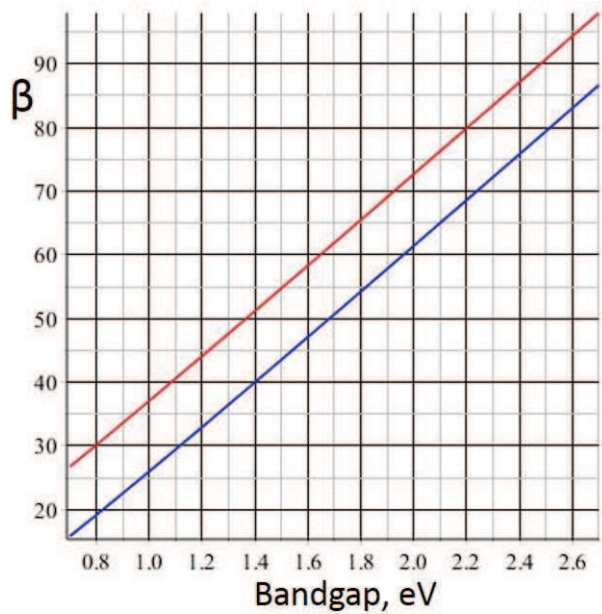}
\caption{The ratio of the photoelectron lifetime to the collection time at optimal PV conversion for the concentrated radiation (red) and unconcentrateted radiation (blue). }
\label{beta}
\end{figure}

Now we will analyze the diffusion processes. For effective photocarrier collection the film thickness, $d$, should be shorter than the corresponding diffusion length related to the carrier collection time,$\sqrt{D\tau_{col}}$ , where $D$ is the diffusion coefficient \cite{12,13}.  Accordig to Eqs. 16 and 18,  $\tau_{col}(V_m) = (4n^2 d\alpha) \tau_R / \beta $. Thus, the requirement of fast photocarrier diffucion to contacts may be presented in the following way,
\begin{equation}
d < \ell_R \sqrt{4n^2 d \alpha \over \beta},
\label{19}
\end{equation}
where $\ell_R$  is the minority carrier diffusion length with respect to emission, $\ell_R = \sqrt{D \tau_R}$ . Finally, requirements of efficient carrier absorption and fast photocarrier diffusion may be presented as 
\begin{equation}
{1 \over 4n^2 \alpha} < d < {4n^2 \ell_R^2 \alpha \over \beta}.
\label{20}
\end{equation}
Let us note that without the reabsorption the last equation would be $\alpha^{-1} <d < \ell_R/\sqrt {\beta} $ (usially $\sqrt {\beta}$ is ignored here \cite{9}), which in turn requires that  $\alpha \ell_R > \sqrt{\beta} \approx 5-7$. This condition cannot be satisfied for any known semiconductor material because of small absorption coefficient near the bandgap ($\alpha$), limited diffusion length of minority carriers, and relatively large factor $\beta$. Typical parameters for Si and GaAs materials used in solar cells are summarized in Table I. 

As seen from Eq. 20, the photon scattering and reabsorption in the semiconductor film improve both the absorption and carrier collection. To approach the SQ limit in PV devices with strong photon reabsorption the corresponding material parameters should satisfy the condition $ \alpha \ell_R > \sqrt{\beta} /(4n^2) \simeq 0.1$, which is fulfilled in good quality Si and GaAs films (see Table I). 

In summary, we obtained analytical solution of SQ model and presented all PV characteristics via well-known Lambert W function (Eqs. 7, 8, 11 - 14). As the Lambert W function is a built-in function in all popular mathematical software, the obtained analytical solution drastically simplifies calculations of  PV  characterisrtics. Key assumption of SQ model, - the chemical equilibrium between photocarriers and emitted photons, - leads to thermodynamic relation between the maximal PV efficiency and output voltage (Eq. 9). The photon trapping introduced by Yablonovich \cite{10} is in fact assumed in the SQ approach, which postulates isotropic equilibrium photon distribution of emitted photons (Eq. 15). Analysis of photoelectron kinetics shows that the carrier collections time at optimal conversion should be substantially shorter than the photocarrier lifetime (Eq. 18 and Fig. 1). The reabsorption increases effective absorption length (Eq. 15) \cite{10} and photocarrier lifetime (Eq. 16). In semiconductor devices, these two factors allows for approaching the SQ limit in semiconductor devices (Eq. 20). At the same time, the long photocarrier lifetime and corresponding long collection time (Eq. 18) may increase nonradiative losses, if, for example, the collection time exceeds the Auger recombination time.Therefore, enhancement of photon-electron coupling via plasmonic modes \cite{14}, whispering gallery modes \cite{15} etc  is more effective way for approaching the SQ efficiencies in semiconductor structures.   

The authors wish to acknowledge Dr. John Little for useful discussions. The work was supported by National Research Council and Army Research Laboratory. 

\begin{table}
\caption{\label{tab:example} Parameters of SI and GaAs PV materials}
\begin{ruledtabular}
\begin{tabular}{llll}
Material  & $\alpha $& $\ell_R$ & $\alpha \cdot \ell_R $ \\
 & cm$^{-1}$& $\mu$m &  \\
 Si & $\sim3$  & $\sim 300$ & $ \sim 0.1 $\\
 GaAs & $\sim 200$  & $ 3-5 $ & $ \sim 0.1$ \\
\end{tabular}
\end{ruledtabular}
\end{table}

\end{document}